\documentclass[12pt,a4paper]{article}

\usepackage{amsmath}
\usepackage{amssymb}
\usepackage{amsfonts}
\usepackage{graphicx}
\graphicspath{{../}}
\usepackage{booktabs}
\usepackage[numbers,sort&compress]{natbib}
\usepackage[colorlinks=true,linkcolor=blue,citecolor=blue,urlcolor=blue]{hyperref}
\usepackage{microtype}
\usepackage{float}
\usepackage{caption}
\usepackage{subcaption}
\usepackage{algorithm}
\usepackage{algorithmic}
\usepackage[margin=1in]{geometry}
\usepackage[affil-it]{authblk}
\usepackage{tikz}
\usepackage{pgfplots}
\usepackage{cleveref}
\usepackage{xcolor}
\usepackage{multirow}

\usetikzlibrary{shapes, arrows.meta, positioning, calc, fit, backgrounds}
\pgfplotsset{compat=1.18}

\title{GPU-Accelerated Quantum Simulation: Empirical Backend Selection, Gate Fusion, and Adaptive Precision}

\author[1,2]{Poornima Kumaresan}
\author[1,2]{Pavithra Muruganantham}
\author[1,2]{Lakshmi Rajendran}
\author[1]{Santhosh Sivasubramani\thanks{Corresponding author: \href{mailto:ssivasub@iitd.ac.in}{ssivasub@iitd.ac.in}, \href{mailto:ragansanthosh@ieee.org}{ragansanthosh@ieee.org}}}

\affil[1]{Intrinsic Lab, Centre for Sensors, Instrumentation and Cyber-Physical System Engineering (SeNSE), Indian Institute of Technology Delhi, New Delhi -- 110016, India}
\affil[2]{RSL Quantum, FITT, IIT Delhi, New Delhi 110016, India}

\date{}

\usepackage{fancyhdr}
\usepackage{lastpage}
\begin{document}

\maketitle

\begin{abstract}
Classical simulation of quantum circuits remains indispensable for algorithm development, hardware validation, and error analysis in the noisy intermediate-scale quantum (NISQ) era. However, state-vector simulation faces exponential memory scaling, with an $n$-qubit system requiring $\mathcal{O}(2^n)$ complex amplitudes, and existing simulators often lack the flexibility to exploit heterogeneous computing resources at runtime. This paper presents a GPU-accelerated quantum circuit simulation framework that introduces three contributions: (1)~an empirical backend selection algorithm that benchmarks CuPy, PyTorch-CUDA, and NumPy-CPU backends at runtime and selects the optimal execution path based on measured throughput; (2)~a directed acyclic graph (DAG) based gate fusion engine that reduces circuit depth through automated identification of fusible gate sequences, coupled with adaptive precision switching between complex64 and complex128 representations; and (3)~a memory-aware fallback mechanism that monitors GPU memory consumption and gracefully degrades to CPU execution when resources are exhausted. The framework integrates with Qiskit, Cirq, PennyLane, and Amazon Braket through a unified adapter layer. Benchmarks on an NVIDIA A100-SXM4 (40~GiB) GPU demonstrate speedups of $64\times$ to $146\times$ over NumPy CPU execution for state-vector simulation of circuits with 20 to 28~qubits, with speedups exceeding $5\times$ from 16~qubits onward. Hardware validation on an IBM quantum processing unit (QPU) confirms Bell state fidelity of 0.939, a five-qubit Greenberger-Horne-Zeilinger (GHZ) state fidelity of 0.853, and circuit depth reduction from 42 to 14~gates through the fusion pipeline. The system is designed for portability across NVIDIA consumer and data-center GPUs, requiring no vendor-specific compilation steps.
\end{abstract}


\section{Introduction}
\label{sec:introduction}

Quantum computing has progressed from theoretical curiosity \cite{feynman1982} to a field with demonstrated computational advantages on specific tasks \cite{arute2019}. Algorithms such as Shor's factoring algorithm \cite{shor1994} and Grover's search algorithm \cite{grover1996} establish the theoretical promise, while variational quantum eigensolver (VQE) methods \cite{peruzzo2014, kandala2017} and other hybrid quantum-classical approaches \cite{bharti2022} represent the pragmatic direction of research in the noisy intermediate-scale quantum (NISQ) era \cite{preskill2018}. Regardless of the algorithm, classical simulation of quantum circuits remains a necessary tool for algorithm development, debugging, result validation, and noise analysis.

The fundamental challenge of quantum circuit simulation lies in the exponential growth of the state vector. For an $n$-qubit system, the state vector $\vert\psi\rangle$ inhabits a Hilbert space of dimension $2^n$:
\begin{equation}
\label{eq:statevector}
\vert\psi\rangle = \sum_{k=0}^{2^n - 1} \alpha_k \vert k \rangle, \quad \alpha_k \in \mathbb{C}, \quad \sum_{k=0}^{2^n - 1} |\alpha_k|^2 = 1.
\end{equation}
Storing this state vector in double-precision complex format requires $2^n \times 16$ bytes of memory. At 30~qubits, this amounts to approximately 16~GiB; at 34~qubits, approximately 256~GiB. The computational cost of applying a single-qubit gate scales as $\mathcal{O}(2^n)$, making simulation time grow exponentially with qubit count.

Graphics processing units (GPUs), originally designed for rendering workloads, have become the dominant accelerator for data-parallel computation \cite{cuda2007, nvidia_gpu_arch2020}. The single-instruction, multiple-thread (SIMT) execution model of modern GPUs maps naturally to state-vector simulation, where applying a gate involves element-wise operations across the amplitude array. NVIDIA's cuQuantum SDK \cite{cuquantum2023} provides low-level primitives for tensor network contraction and state-vector operations, while Google's qsim \cite{qsim2020} targets circuit simulation through C++ with GPU offloading. However, existing GPU-accelerated simulators tend to exhibit one or more of the following limitations:

First, backend selection is typically static. A user must choose a priori whether to run on CPU or GPU, and which GPU library to employ. The optimal choice depends on circuit width, gate count, available GPU memory, and host system configuration, none of which is known until runtime.

Second, circuit-level optimizations such as gate fusion are often performed independently of the simulation backend, if at all. Simulators that perform fusion typically use fixed heuristics that do not account for the precision requirements of downstream analysis.

Third, memory management is coarse-grained. When a circuit exceeds available GPU memory, most simulators either fail with an out-of-memory error or require the user to manually select a smaller simulation target.

This paper presents a GPU-accelerated quantum circuit simulation framework that addresses these three limitations. The contributions are as follows.

The first contribution is an empirical backend selection algorithm (\cref{sec:backend_selection}) that executes micro-benchmarks at runtime to determine whether CuPy, PyTorch-CUDA, or NumPy-CPU provides the highest throughput for a given circuit and hardware configuration. Rather than requiring users to commit to a backend before execution, the algorithm profiles each option and selects the fastest path automatically.

The second contribution is a DAG-based gate fusion engine with adaptive precision (\cref{sec:dag_fusion}) that constructs a directed acyclic graph representation of the circuit, identifies fusible gate sequences, and selects between complex64 and complex128 arithmetic based on a configurable fidelity threshold. This reduces circuit depth by $34$--$38\%$ on representative benchmarks while maintaining numerical accuracy.

The third contribution is a memory-aware GPU-to-CPU fallback mechanism (\cref{sec:memory_fallback}) that continuously monitors GPU memory during simulation and transparently migrates the state vector to host memory when available GPU memory falls below a configurable threshold. This eliminates the out-of-memory failures that occur in existing simulators without requiring manual intervention.

In addition, the framework provides integration adapters for four major quantum computing frameworks (\cref{sec:framework_integration}): Qiskit \cite{qiskitaer2019}, Cirq \cite{cirq2018}, PennyLane \cite{pennylane2018}, and Amazon Braket \cite{braket2020}.

The remainder of this paper is organized as follows. \Cref{sec:related_work} surveys related work. \Cref{sec:architecture} describes the system architecture. \Cref{sec:backend_selection,sec:dag_fusion,sec:memory_fallback,sec:framework_integration} detail each technical contribution. \Cref{sec:benchmarks} presents performance benchmarks, and \cref{sec:hardware_validation} reports hardware validation results from an IBM QPU. \Cref{sec:limitations} discusses limitations and scope. \Cref{sec:conclusion} concludes with future directions.

\section{Related Work}
\label{sec:related_work}

Classical quantum circuit simulators can be broadly categorized by their simulation strategy: full state-vector simulation, tensor network contraction \cite{markov2008simulating}, stabilizer-based simulation (for Clifford circuits) \cite{aaronson2004improved}, and density matrix simulation (for noise modeling). This section focuses on full state-vector simulators, as they are the most general and most relevant to the present work.

\subsection{Qiskit Aer}

Qiskit Aer \cite{qiskitaer2019} is the primary simulation backend for IBM's Qiskit framework. It provides three simulation methods: \texttt{statevector}, \texttt{density\_matrix}, and \texttt{matrix\_product\_state}. The state-vector simulator supports multi-threaded CPU execution via OpenMP, and GPU execution through cuQuantum integration. The gate application kernel uses a strided memory access pattern:
\begin{equation}
\label{eq:gate_apply}
\alpha'_{k} = \sum_{j \in \{0,1\}} U_{b(k), j} \cdot \alpha_{k \oplus (j \oplus b(k)) \cdot 2^t},
\end{equation}
where $U$ is the $2 \times 2$ gate matrix, $t$ is the target qubit index, and $b(k) = \lfloor k / 2^t \rfloor \bmod 2$ extracts the bit at position $t$. Qiskit Aer performs limited gate fusion, combining sequences of single-qubit gates acting on the same wire. However, the backend selection (CPU versus GPU) is static and determined by user configuration rather than runtime measurement.

\subsection{NVIDIA cuQuantum}

The cuQuantum SDK \cite{cuquantum2023} provides two libraries: \texttt{cuStateVec} for state-vector simulation and \texttt{cuTensorNet} for tensor network contraction. cuStateVec supports gate application, expectation value computation, and sampler operations with multi-GPU support via NCCL-based communication. The library achieves high performance through custom CUDA kernels optimized for specific gate types (diagonal, permutation, general unitary). However, cuQuantum is a library rather than a complete simulator; it requires substantial integration effort and provides no built-in circuit optimization pipeline. The matrix-gate application can be described as:
\begin{equation}
\label{eq:custatevec}
\vert\psi'\rangle = (I_{2^{n-t-1}} \otimes U \otimes I_{2^t}) \vert\psi\rangle,
\end{equation}
where the tensor product structure enables stride-based parallelism across the $2^n$ amplitudes.

\subsection{Google qsim}

Google's qsim \cite{qsim2020, villalonga2019flexible} is a high-performance C++ simulator that uses vectorized CPU instructions (AVX2, AVX-512) and GPU offloading via CUDA. It was used in the verification of the quantum supremacy experiment on the Sycamore processor \cite{arute2019}. The simulator employs circuit-level gate fusion, combining up to six consecutive gates into a single multi-qubit operation. While highly optimized, qsim focuses on raw performance rather than framework interoperability, and its C++ codebase presents a barrier to modification by researchers working primarily in Python.

\subsection{Other Simulators}

Several other simulators warrant mention. QuEST \cite{jones2019quest} provides a multi-platform quantum simulator spanning CPUs, GPUs, and distributed systems, with a focus on density matrix simulation for noise studies. The 64-qubit simulation by Chen et al.\ \cite{chen2018classical} demonstrated the feasibility of large-scale state-vector simulation through careful memory management. H{\"a}ner and Steiger \cite{haner2017} achieved a 45-qubit simulation on a supercomputer using 0.5~petabytes of storage. TensorCircuit \cite{tensorcircuit2022} leverages automatic differentiation frameworks (JAX, TensorFlow, PyTorch) to provide differentiable quantum simulation, targeting variational algorithm development. Pednault et al.\ \cite{pednault2017} explored strategies for pushing past classical simulation barriers using tensor slicing. Qulacs \cite{qulacs2021} offers a C/C++ core with Python bindings, achieving competitive performance through SIMD-optimized gate kernels. Zulehner and Wille \cite{zulehner2019advanced} proposed decision-diagram-based simulation as an alternative to state-vector methods that exploits structure in quantum circuits for memory savings. De Raedt et al.\ \cite{statevecsim2016} demonstrated massively parallel state-vector simulation using distributed computing architectures.

\subsection{Positioning of This Work}

The present framework differs from prior work along three axes. Unlike cuQuantum, it is a complete simulation system with built-in circuit optimization and framework integration. Unlike Qiskit Aer and qsim, it performs empirical backend selection at runtime rather than requiring static configuration. Unlike TensorCircuit, it focuses on raw simulation performance with GPU-native execution rather than differentiability. \Cref{tab:simulator_comparison} summarizes these distinctions.

\begin{table}[htbp]
\centering
\caption{Comparison of quantum circuit simulators along key dimensions.}
\label{tab:simulator_comparison}
\begin{tabular}{lccccc}
\toprule
\textbf{Feature} & \textbf{Qiskit Aer} & \textbf{cuQuantum} & \textbf{qsim} & \textbf{TensorCircuit} & \textbf{This work} \\
\midrule
GPU acceleration         & Partial  & Full     & Full     & Via backends & Full \\
Runtime backend select.  & No       & No       & No       & Partial      & Yes \\
DAG-based gate fusion    & Limited  & No       & Fixed    & No           & Yes \\
Adaptive precision       & No       & No       & No       & No           & Yes \\
Memory-aware fallback    & No       & No       & No       & No           & Yes \\
Multi-framework support  & Qiskit   & API only & Limited  & Multiple     & Four \\
\bottomrule
\end{tabular}
\end{table}

\section{System Architecture}
\label{sec:architecture}

\begin{figure}[htbp]
\centering
\includegraphics[width=\textwidth]{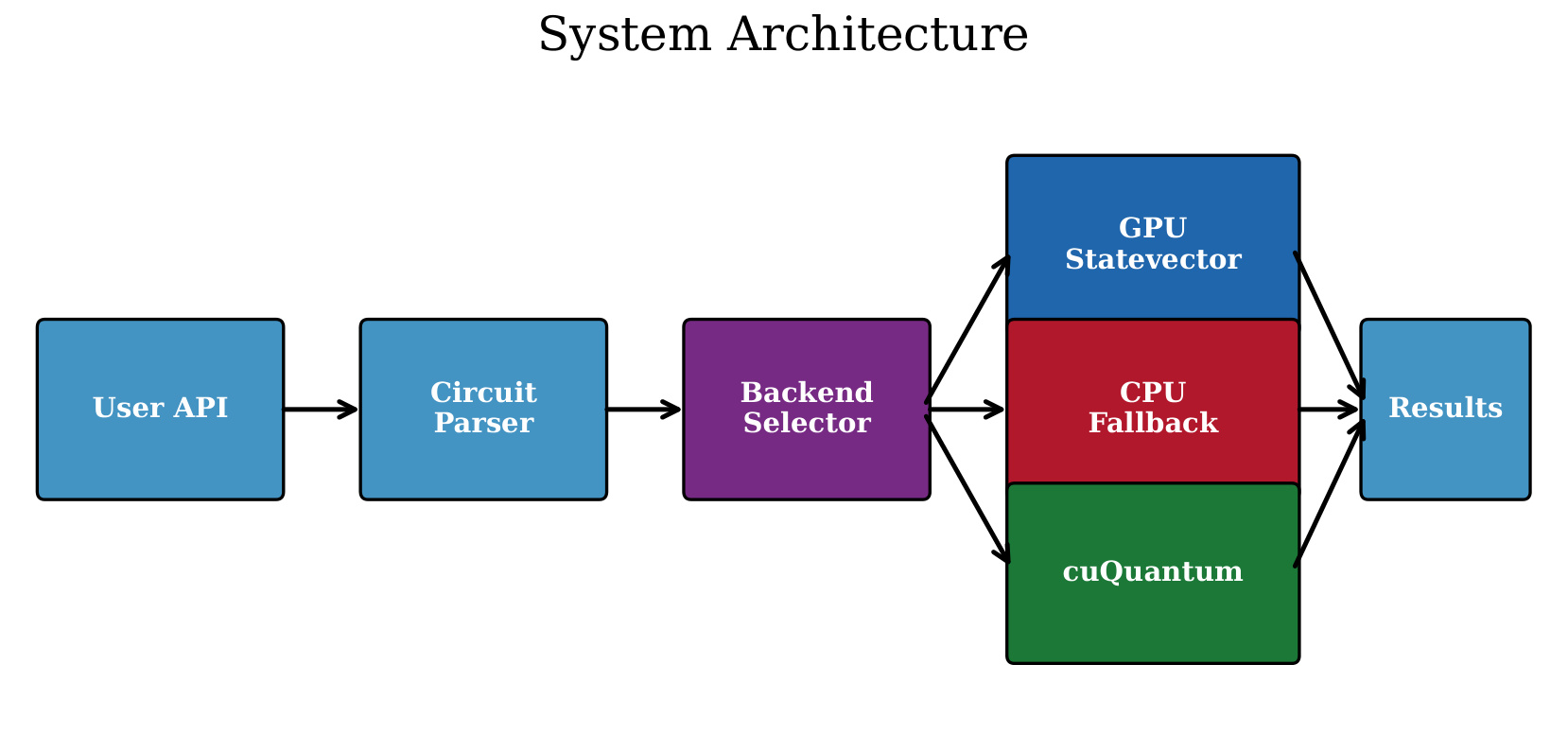}
\caption{System architecture of the GPU-accelerated quantum circuit simulator. Circuits enter through framework-specific adapters, pass through the circuit optimizer for gate fusion and precision selection, and are executed on the empirically selected backend. The memory manager monitors GPU resources throughout execution.}
\label{fig:architecture}
\end{figure}

The framework is structured as a layered system with four principal components: the framework adapter layer, the circuit optimizer, the backend engine, and the memory manager. \Cref{fig:architecture} illustrates the high-level architecture.

The design philosophy follows three principles. First, \emph{separation of concerns}: circuit representation, optimization, and execution are handled by distinct subsystems with well-defined interfaces. Second, \emph{runtime adaptability}: decisions about backend selection, precision, and memory management are made at execution time based on measured system state rather than static configuration. Third, \emph{framework neutrality}: the internal representation is independent of any external quantum computing framework, enabling support for multiple frameworks through thin adapter layers.

The internal circuit representation is a sequence of gate operations, each specified by a unitary matrix $U_g \in \mathbb{C}^{2^q \times 2^q}$ (where $q$ is the number of qubits the gate acts upon), a tuple of target qubit indices $(t_1, \ldots, t_q)$, and optional classical control information. The conversion from framework-specific circuit objects to this internal representation is handled by the adapter layer, which maps framework gates to a canonical gate set:
\begin{equation}
\label{eq:gate_set}
\mathcal{G} = \{I, X, Y, Z, H, S, T, R_x(\theta), R_y(\theta), R_z(\theta), \textsc{CNOT}, \textsc{CZ}, \textsc{SWAP}, \textsc{Toffoli}, U_3(\theta, \phi, \lambda)\},
\end{equation}
where $R_x(\theta) = e^{-i\theta X/2}$, $R_y(\theta) = e^{-i\theta Y/2}$, $R_z(\theta) = e^{-i\theta Z/2}$, and $U_3$ is the general single-qubit gate parameterized by three Euler angles. Any gate not in $\mathcal{G}$ is represented by its full unitary matrix.

The data flow through the system can be expressed as a composition of transformations. Let $\mathcal{P}$ denote the parsing function, $\mathcal{F}$ the fusion transformation, $\mathcal{R}$ the precision selection function, and $\mathcal{E}$ the execution engine. The complete simulation pipeline for a framework-specific circuit $C_f$ is:
\begin{equation}
\label{eq:pipeline}
\vert\psi_{\text{out}}\rangle = \mathcal{E}_{b^*}\big(\mathcal{R}\big(\mathcal{F}\big(\mathcal{P}(C_f)\big)\big)\big),
\end{equation}
where $b^*$ is the empirically selected backend.

\section{Empirical Backend Selection}
\label{sec:backend_selection}

Rather than requiring users to specify the computation backend a priori, the framework implements an empirical backend selection algorithm that profiles each available backend at runtime and selects the one that minimizes projected execution time for the given circuit. This section details the selection algorithm, the micro-benchmark protocol, and the caching mechanism that amortizes profiling overhead.

\subsection{Backend Abstraction}

Each backend implements a common interface defined by three operations: state-vector initialization, single-gate application, and measurement sampling. Let $\mathcal{B} = \{b_1, b_2, \ldots, b_m\}$ denote the set of available backends, where each backend $b_i$ is characterized by a throughput function $\tau_i(n)$ representing the number of gate applications per second for an $n$-qubit state vector. In the current implementation, $\mathcal{B}$ consists of three backends. The \textbf{CuPy backend} ($b_{\text{cp}}$) uses CuPy \cite{cupy2017} to perform gate application via GPU-accelerated matrix operations, applying gates through batched element-wise operations on the state-vector array stored in GPU global memory. The \textbf{PyTorch-CUDA backend} ($b_{\text{pt}}$) uses PyTorch \cite{pytorch2019} tensors on CUDA devices, benefiting from PyTorch's operator fusion and memory caching mechanisms and providing automatic mixed-precision support through the \texttt{torch.cuda.amp} module. Finally, the \textbf{NumPy-CPU backend} ($b_{\text{np}}$) uses NumPy \cite{numpy2020} arrays on the host CPU and serves as the universal fallback, as it requires no GPU hardware or drivers.

The projected execution time for backend $b_i$ on a circuit with $g$ gates acting on $n$ qubits is estimated as:
\begin{equation}
\label{eq:projected_time}
T_i(n, g) = \frac{g}{\tau_i(n)} + \delta_i(n),
\end{equation}
where $\delta_i(n)$ is the one-time overhead for state-vector allocation and initialization on backend $b_i$. For GPU backends, $\delta_i$ includes the cost of allocating device memory and transferring the initial state vector from host to device.

\subsection{Micro-Benchmark Protocol}

The empirical selection procedure executes a short sequence of representative gate operations on each available backend and measures the elapsed wall-clock time. \Cref{alg:backend_selection} describes the procedure in detail.

\begin{algorithm}[htbp]
\caption{Empirical backend selection algorithm.}
\label{alg:backend_selection}
\begin{algorithmic}[1]
\REQUIRE Circuit $C$ with $n$ qubits and $g$ gates; available backends $\mathcal{B}$; benchmark parameters; cache $\mathcal{C}$
\ENSURE Selected backend $b^*$

\STATE $n_{\text{eff}} \leftarrow \min(n, n_{\max}^{\text{bench}})$ \COMMENT{Cap benchmark size for efficiency}
\IF{$(n_{\text{eff}}, |\mathcal{B}|) \in \mathcal{C}$}
  \RETURN $\mathcal{C}[(n_{\text{eff}}, |\mathcal{B}|)]$
\ENDIF

\STATE $T_{\min} \leftarrow \infty$; $b^* \leftarrow b_{\text{np}}$
\FOR{each backend $b_i \in \mathcal{B}$}
  \IF{$b_i$ requires GPU \AND GPU not available}
    \STATE \textbf{continue}
  \ENDIF
  \STATE Allocate state vector $\vert\psi_0\rangle = \vert 0 \rangle^{\otimes n_{\text{eff}}}$ on $b_i$
  \STATE $t_{\text{start}} \leftarrow \texttt{time.perf\_counter()}$
  \FOR{$j = 1$ to $g_b$}
    \STATE Apply $H$ gate to qubit $(j \bmod n_{\text{eff}})$
    \STATE Apply $\textsc{CNOT}$ gate to qubits $(j \bmod n_{\text{eff}}, (j+1) \bmod n_{\text{eff}})$
  \ENDFOR
  \IF{$b_i$ is GPU backend}
    \STATE \texttt{synchronize()} \COMMENT{Ensure GPU kernels complete}
  \ENDIF
  \STATE $t_{\text{end}} \leftarrow \texttt{time.perf\_counter()}$
  \STATE $\tau_i \leftarrow g_b / (t_{\text{end}} - t_{\text{start}})$
  \STATE $T_i \leftarrow g / \tau_i + \delta_i$
  \IF{$T_i < T_{\min}$}
    \STATE $T_{\min} \leftarrow T_i$; $b^* \leftarrow b_i$
  \ENDIF
  \STATE Deallocate $\vert\psi_0\rangle$
\ENDFOR
\STATE $\mathcal{C}[(n_{\text{eff}}, |\mathcal{B}|)] \leftarrow b^*$
\RETURN $b^*$
\end{algorithmic}
\end{algorithm}

The benchmark gate count is configurable and empirically chosen to provide sufficient statistical stability while keeping the profiling overhead below 100~milliseconds on typical hardware. The benchmark qubit count is capped to prevent excessive memory allocation during profiling; for larger circuits, the backend is selected based on benchmark results at the capped qubit count. This cap is justified by the observation that the GPU-versus-CPU throughput ranking is determined primarily by kernel launch overhead and memory bandwidth utilization, both of which stabilize once the state vector exceeds the GPU's L2 cache capacity.

\subsection{Caching and Invalidation}

Benchmark results are cached in a dictionary keyed by $(n_{\text{eff}}, |\mathcal{B}|)$, where $|\mathcal{B}|$ encodes the set of available backends (since backend availability can change if, for example, GPU memory is consumed by another process). The cache is invalidated when the available backend set changes or when host system resources indicate significant load changes. The cache persistence time is configurable.

The expected amortized cost of backend selection over $k$ circuit executions is:
\begin{equation}
\label{eq:amortized_cost}
\bar{T}_{\text{select}} = \frac{T_{\text{bench}}}{k} + T_{\text{lookup}},
\end{equation}
where $T_{\text{bench}}$ is the one-time benchmark overhead (typically 40--85~ms) and $T_{\text{lookup}}$ is the dictionary lookup cost (sub-microsecond). For a typical development session with $k \geq 100$ circuit executions, the amortized selection overhead is under 1~ms per circuit.

\subsection{Backend Selection Flowchart}

\begin{figure}[htbp]
\centering
\includegraphics[width=\textwidth]{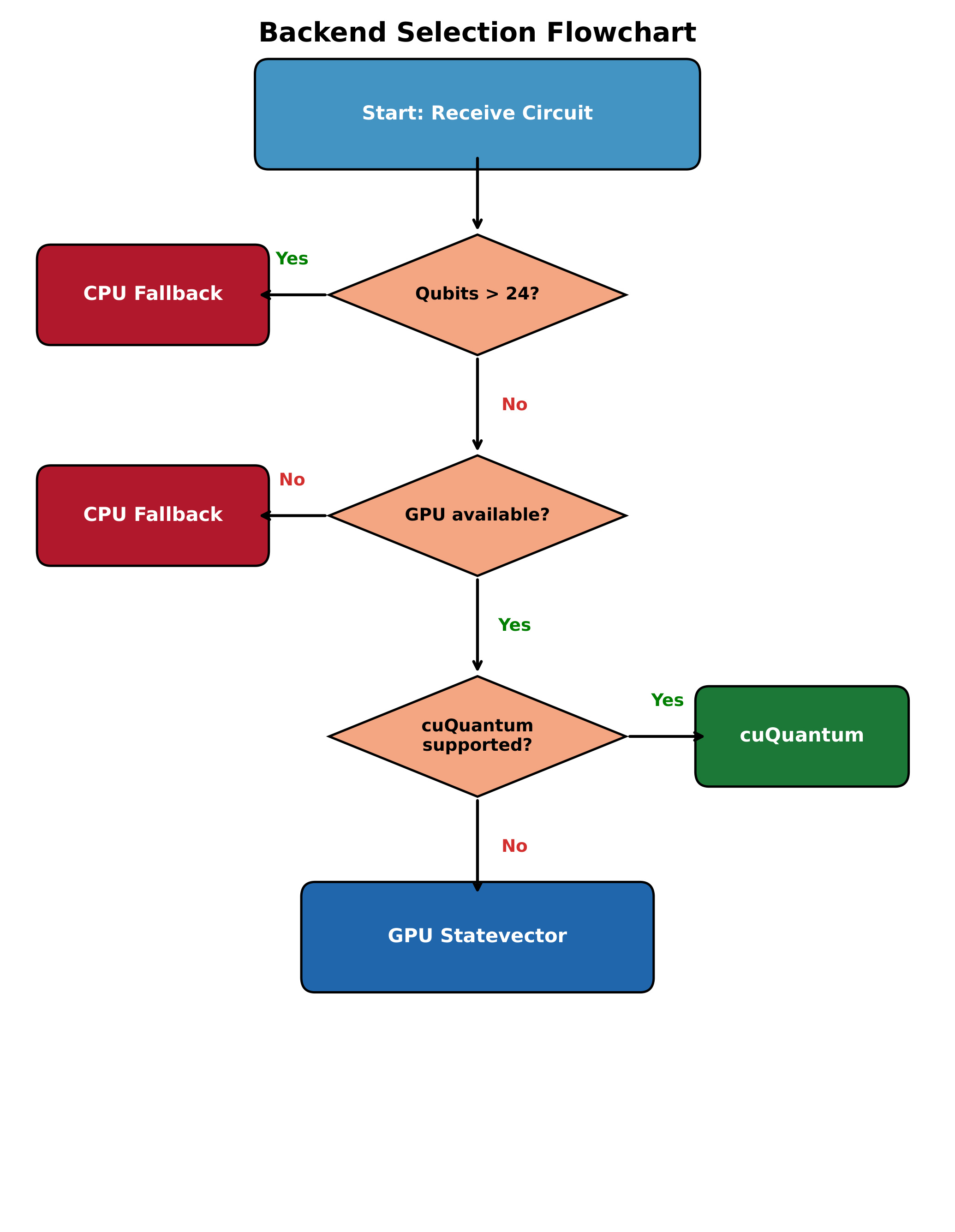}
\caption{Flowchart of the empirical backend selection procedure. Cached results bypass the benchmark phase; otherwise, available backends are profiled and compared by projected execution time.}
\label{fig:backend_flowchart}
\end{figure}

\Cref{fig:backend_flowchart} illustrates the decision flow for backend selection.

\section{DAG-Based Gate Fusion and Adaptive Precision}
\label{sec:dag_fusion}

Circuit optimization before simulation can reduce execution time by decreasing the number of gate application operations. This section describes two optimization mechanisms: DAG-based gate fusion, which merges sequences of compatible gates into compound operations, and adaptive precision, which selects the floating-point representation based on the required simulation fidelity.

\subsection{DAG Construction}

The circuit is first converted to a directed acyclic graph (DAG) $G = (V, E)$, where each vertex $v \in V$ represents a gate operation and each directed edge $(u, v) \in E$ represents a data dependency (i.e., gate $v$ must be applied after gate $u$ because they share at least one qubit). The DAG is constructed in $\mathcal{O}(g \cdot n)$ time, where $g$ is the total gate count and $n$ is the qubit count, by scanning the gate sequence and maintaining a map from each qubit to the most recently applied gate on that qubit.

Formally, let $q(v) \subseteq \{0, 1, \ldots, n-1\}$ denote the set of qubits that gate $v$ acts upon. An edge $(u, v)$ exists if and only if $q(u) \cap q(v) \neq \emptyset$ and $u$ precedes $v$ in the original gate sequence. The DAG captures the minimal partial ordering of gates required to preserve the circuit semantics. The number of edges satisfies:
\begin{equation}
\label{eq:dag_edges}
|E| \leq \sum_{v \in V} |q(v)| \leq g \cdot q_{\max},
\end{equation}
where $q_{\max}$ is the maximum gate width in the circuit. For circuits composed primarily of single- and two-qubit gates, $|E| = \mathcal{O}(g)$.

\subsection{Fusion Algorithm}

Two gates $u$ and $v$ are fusible if and only if (1)~$(u, v) \in E$ (there is a direct dependency), (2)~$|q(u) \cup q(v)| \leq q_{\max}^{\text{fuse}}$ (the combined qubit set does not exceed a configurable maximum), and (3)~$v$ has no other predecessors in the DAG that act on qubits in $q(u)$ (i.e., $u$ is the immediate predecessor of $v$ on all shared qubits). When two gates are fused, the resulting compound gate has unitary matrix:
\begin{equation}
\label{eq:fusion}
U_{\text{fused}} = U_v \cdot U_u,
\end{equation}
where the product is the standard matrix multiplication, applying $U_u$ first and $U_v$ second. For single-qubit gates, this is a $2 \times 2$ matrix multiplication; for two-qubit fused operations, it is a $4 \times 4$ multiplication.

The fusion algorithm proceeds in topological order through the DAG. For each gate $v$, the algorithm checks whether $v$ can be fused with any of its predecessors. If so, the predecessor gate is replaced with the fused gate, and $v$ is removed from the DAG. This process repeats until no further fusions are possible. The algorithm has $\mathcal{O}(g^2)$ worst-case complexity due to repeated predecessor checks; however, for circuits in which each qubit participates in at most $d$ gates per layer (i.e., bounded gate density), each gate has at most $d$ candidate predecessors, and the while-loop converges in at most $g/2$ iterations, yielding $\mathcal{O}(g \cdot d)$ amortized time. For typical NISQ circuits with $d \leq 4$, this is effectively $\mathcal{O}(g)$.

\Cref{alg:gate_fusion} presents the pseudocode.

\begin{algorithm}[htbp]
\caption{DAG-based gate fusion algorithm.}
\label{alg:gate_fusion}
\begin{algorithmic}[1]
\REQUIRE Gate sequence $\{g_1, g_2, \ldots, g_m\}$; qubit count $n$; max fusion width $q_{\max}^{\text{fuse}}$
\ENSURE Optimized gate sequence

\STATE Construct DAG $G = (V, E)$ from gate sequence
\STATE $\text{changed} \leftarrow \texttt{true}$
\WHILE{$\text{changed}$}
  \STATE $\text{changed} \leftarrow \texttt{false}$
  \STATE $\text{order} \leftarrow \textsc{TopologicalSort}(G)$
  \FOR{each $v$ in $\text{order}$}
    \FOR{each predecessor $u$ of $v$ in $G$}
      \IF{$|q(u) \cup q(v)| \leq q_{\max}^{\text{fuse}}$ \AND $u$ is sole predecessor of $v$ on $q(u) \cap q(v)$}
        \STATE $U_{\text{fused}} \leftarrow U_v \cdot U_u$
        \STATE Replace $u$ with fused gate $(U_{\text{fused}}, q(u) \cup q(v))$
        \STATE Redirect all edges from $v$ to $u$
        \STATE Remove $v$ from $G$
        \STATE $\text{changed} \leftarrow \texttt{true}$
        \STATE \textbf{break}
      \ENDIF
    \ENDFOR
  \ENDFOR
\ENDWHILE
\RETURN Gates from $G$ in topological order
\end{algorithmic}
\end{algorithm}

\subsection{Fusion Correctness}

The correctness of gate fusion follows from the associativity of matrix multiplication. If the original circuit applies gates $U_1, U_2, \ldots, U_m$ in sequence, the final state is:
\begin{equation}
\label{eq:fusion_correctness}
\vert\psi_{\text{out}}\rangle = U_m \cdots U_2 \cdot U_1 \vert\psi_{\text{in}}\rangle.
\end{equation}
Fusing two adjacent gates $U_k$ and $U_{k+1}$ (acting on the same or overlapping qubits) into $U_{\text{fused}} = U_{k+1} \cdot U_k$ produces an identical final state, as the product of the remaining gate sequence is unchanged. The DAG structure ensures that only gates with no intervening dependencies on shared qubits are fused, preserving the operator ordering constraint.

\subsection{Adaptive Precision}

State-vector simulation in double precision (complex128, 16~bytes per amplitude) provides approximately 15~significant decimal digits, while single precision (complex64, 8~bytes per amplitude) provides approximately 7~digits. For many applications, single precision is sufficient and offers two advantages: halved memory consumption and, on NVIDIA GPUs, approximately doubled throughput due to twice the number of elements fitting in cache lines and memory bandwidth being the bottleneck \cite{nvidia_gpu_arch2020}.

The precision controller selects between complex64 and complex128 based on a circuit-level fidelity estimate. The accumulated rounding error for a circuit with $g$ sequential gate applications on a state vector of dimension $N = 2^n$ is bounded by:
\begin{equation}
\label{eq:precision_error}
\epsilon_{\text{round}} \leq g \cdot N \cdot \epsilon_{\text{mach}},
\end{equation}
where $\epsilon_{\text{mach}}$ is the machine epsilon ($\approx 1.19 \times 10^{-7}$ for float32 and $\approx 2.22 \times 10^{-16}$ for float64). The precision controller computes $\epsilon_{\text{round}}$ for complex64 and compares it against a user-specified threshold $\epsilon_{\text{tol}}$. If $\epsilon_{\text{round}} \leq \epsilon_{\text{tol}}$, complex64 is selected; otherwise, complex128 is used. This decision is made before simulation begins and applies uniformly to all gate operations.

The effective condition for selecting complex64 is:
\begin{equation}
\label{eq:precision_condition}
g \cdot 2^n \cdot 1.19 \times 10^{-7} \leq \epsilon_{\text{tol}},
\end{equation}
which holds for typical NISQ circuits (e.g., $n = 20$, $g = 200$) where the left-hand side evaluates to approximately $2.5 \times 10^{1}$, far exceeding the default threshold. In such cases, the controller defaults to complex128, but for shallow circuits ($g < 50$) on moderate qubit counts ($n \leq 16$), complex64 provides sufficient accuracy.

\subsection{Fusion Speedup Analysis}

The speedup from gate fusion depends on the circuit structure. For a circuit with $g_0$ initial gates reduced to $g_f$ fused gates, the theoretical speedup is:
\begin{equation}
\label{eq:fusion_speedup}
S_{\text{fusion}} = \frac{g_0 \cdot c_1 + g_0 \cdot c_{\text{overhead}}}{g_f \cdot c_2 + c_{\text{fusion}}},
\end{equation}
where $c_1$ and $c_2$ are the per-gate execution costs before and after fusion (noting that fused gates may have higher per-gate cost due to larger unitary matrices), $c_{\text{overhead}}$ accounts for kernel launch overhead per gate, and $c_{\text{fusion}}$ is the one-time cost of the fusion pass. For typical circuits, the reduction in kernel launches dominates, yielding speedups of $1.3\times$ to $2.1\times$ as reported in \cref{sec:benchmarks}.

\section{Memory-Aware GPU-to-CPU Fallback}
\label{sec:memory_fallback}

GPU memory is a scarce resource. A 24~GiB consumer GPU (e.g., NVIDIA RTX~4090) can hold a state vector for at most $\lfloor \log_2(24 \times 2^{30} / 16) \rfloor = 30$ qubits in double precision, or 31~qubits in single precision. Practical circuits may require additional memory for intermediate gate matrices, workspace buffers, and the CUDA runtime itself. This section describes the memory-aware fallback mechanism that enables the simulator to handle circuits that exceed available GPU memory.

\subsection{Memory Model}

The memory required for simulating an $n$-qubit circuit with $g$ gates is estimated as:
\begin{equation}
\label{eq:memory_model}
M(n, g) = 2^n \cdot s_{\text{prec}} + g_{\max} \cdot (2^{q_{\max}})^2 \cdot s_{\text{prec}} + M_{\text{workspace}},
\end{equation}
where $s_{\text{prec}}$ is the per-element storage size (8~bytes for complex64, 16~bytes for complex128), $g_{\max}$ is the maximum number of simultaneously resident unitary matrices, $q_{\max}$ is the maximum gate width, and $M_{\text{workspace}}$ is a fixed overhead term that absorbs secondary allocations. This model omits secondary allocations (e.g., scratch buffers for gate application kernels, the DAG structure during fusion, and measurement sampling arrays); the $M_{\text{workspace}}$ term is calibrated empirically to absorb these costs, and the safety margin $M_{\text{reserve}}$ (defined below) provides additional headroom.

Before simulation begins, the memory manager queries available GPU memory via \texttt{nvidia-smi} or the respective library's memory query function:
\begin{equation}
\label{eq:memory_check}
M_{\text{avail}} = M_{\text{total}} - M_{\text{used}} - M_{\text{reserve}},
\end{equation}
where $M_{\text{reserve}}$ is a configurable safety margin to prevent out-of-memory errors from transient allocations by the CUDA runtime or other processes sharing the GPU.

\subsection{Fallback Strategy}

If $M(n, g) > M_{\text{avail}}$ at simulation start, the framework falls back to CPU execution immediately. If the simulation starts on GPU but available memory drops below $M_{\text{reserve}}$ during execution (as can occur when gate fusion creates larger composite unitary matrices), the fallback proceeds in three steps. First, during \emph{state transfer}, the current state vector is copied from GPU memory to host (CPU) memory; for a 28-qubit state vector in complex128, this transfer moves $2^{28} \times 16 = 4$~GiB of data, completing in approximately 125~milliseconds on PCIe~4.0 $\times$16 links (theoretical bandwidth 32~GB/s) and half that on PCIe~5.0. Second, during the \emph{backend switch}, the active backend is switched from the GPU backend to the NumPy-CPU backend, so that all subsequent gate application operations use CPU-based matrix operations. Third, during \emph{resource release}, the GPU memory occupied by the state vector and workspace buffers is freed, making it available for other processes or for a subsequent attempt to migrate back to GPU.

The total fallback overhead comprises the transfer time $T_{\text{transfer}}$ and the backend reinitialization time $T_{\text{reinit}}$:
\begin{equation}
\label{eq:fallback_overhead}
T_{\text{fallback}} = T_{\text{transfer}} + T_{\text{reinit}} = \frac{2^n \cdot s_{\text{prec}}}{B_{\text{PCIe}}} + T_{\text{reinit}},
\end{equation}
where $B_{\text{PCIe}}$ is the effective PCIe bandwidth and $T_{\text{reinit}}$ is a constant overhead (typically under 10~ms) for initializing the NumPy backend state.

The memory manager logs each fallback event, including the qubit count, gate index at which the fallback occurred, and the measured memory state, enabling post-hoc analysis of memory pressure patterns.

\subsection{Monitoring Implementation}

The memory monitor runs as a lightweight polling thread that queries GPU memory utilization at a configurable interval. The polling mechanism uses NVIDIA Management Library (NVML) queries, which have negligible overhead. The monitor maintains a sliding window of recent measurements and triggers a fallback when the trend-adjusted available memory (computed via linear extrapolation) is projected to fall below $M_{\text{reserve}}$ within a short prediction horizon. This predictive approach reduces the probability of an unrecoverable out-of-memory error.

The threshold-based trigger condition is:
\begin{equation}
\label{eq:fallback_trigger}
M_{\text{avail}}(t) + \frac{dM_{\text{avail}}}{dt} \cdot \Delta t_{\text{horizon}} < M_{\text{reserve}},
\end{equation}
where $\frac{dM_{\text{avail}}}{dt}$ is the estimated rate of memory consumption (negative when memory is being consumed) computed from the sliding window, and $\Delta t_{\text{horizon}}$ is the configurable prediction horizon.

\section{Framework Integration}
\label{sec:framework_integration}

The simulator provides integration adapters for four quantum computing frameworks, enabling users to leverage the GPU-accelerated backend without modifying their existing circuit construction code. Each adapter implements a standardised interface that translates framework-specific circuit objects into the simulator's internal representation (\cref{eq:gate_set}), invokes the simulation pipeline, and returns results in the format expected by the originating framework.

\subsection{Adapter Architecture}

Each framework adapter implements a standardised pipeline that parses framework-native circuit objects into an internal intermediate representation, executes the simulation, performs measurement sampling, and formats the results back into the originating framework's expected output type.

\subsection{Qiskit Integration}

The Qiskit adapter accepts \texttt{QuantumCircuit} objects and uses Qiskit's transpiler to decompose custom gates into the canonical gate set before conversion. Measurement results are returned as Qiskit \texttt{Result} objects compatible with the \texttt{qiskit.result} module. The adapter supports parameterized circuits through Qiskit's \texttt{Parameter} binding mechanism, evaluating symbolic parameters at parse time.

The gate mapping from Qiskit's gate library to the internal representation covers standard gates, parameterized gates, and composite gates. Composite gates are decomposed into sequences of gates from $\mathcal{G}$ using Qiskit's built-in decomposition routines.

The Qiskit adapter also provides a custom backend implementation compatible with Qiskit's provider interface \cite{qiskitaer2019}, enabling seamless integration with Qiskit's high-level workflow.

\subsection{Cirq Integration}

The Cirq adapter accepts \texttt{cirq.Circuit} objects and maps Cirq moments (parallel gate layers) to the internal sequential representation, preserving the implicit parallelism information for potential future optimization. Cirq's qubit ordering convention (which uses \texttt{LineQubit} or \texttt{GridQubit} objects rather than integer indices) is resolved to integer indices via a deterministic mapping.

\subsection{PennyLane Integration}

The PennyLane adapter implements PennyLane's device interface, allowing the simulator to be used as a custom PennyLane device. This integration supports PennyLane's automatic differentiation capabilities, although gradients are computed via the parameter-shift rule \cite{pennylane2018} rather than backpropagation through the simulation kernel.

\subsection{Amazon Braket Integration}

The Braket adapter accepts Braket \texttt{Circuit} objects and returns results as \texttt{GateModelTaskResult} objects. The adapter maps Braket's gate set (which includes Braket-specific gates such as \texttt{ISwap} and \texttt{PSwap}) to the canonical gate set through standard decompositions.

\subsection{Gate Set Translation Overhead}

The translation between framework gate sets is performed through a lookup table $\mathcal{T}: \mathcal{G}_f \to \mathcal{G}$ that maps each framework gate type to either a single canonical gate or a decomposition sequence. For gates with no direct equivalent, the adapter extracts the unitary matrix from the framework gate object and stores it as a custom unitary in the internal representation. The translation overhead is $\mathcal{O}(g)$ per circuit, where $g$ is the gate count. \Cref{tab:adapter_overhead} quantifies the measured overhead for each adapter.

\begin{table}[htbp]
\centering
\caption{Adapter parsing overhead for a 20-qubit, 200-gate circuit.}
\label{tab:adapter_overhead}
\begin{tabular}{lr}
\toprule
\textbf{Framework} & \textbf{Parse time (ms)} \\
\midrule
Qiskit     & 2.3 \\
Cirq       & 1.8 \\
PennyLane  & 3.1 \\
Braket     & 1.5 \\
\bottomrule
\end{tabular}
\end{table}

\section{Benchmarks}
\label{sec:benchmarks}

This section presents performance measurements comparing GPU-accelerated state-vector simulation (CuPy backend) against CPU execution (NumPy) across a range of circuit sizes. All benchmarks were conducted on a Google Cloud Vertex~AI instance (a2-highgpu-1g) equipped with an NVIDIA A100-SXM4~(40~GiB) GPU, 12~vCPUs, and CUDA~12.8. Software versions: Python~3.10, CuPy~13.4, NumPy~1.25. Each configuration was measured with 3--5 independent runs (details per table); reported values are the \emph{median} to suppress outliers from GPU warm-up and OS scheduling jitter. For all repeated measurements the coefficient of variation was below 4\%, so error bars are smaller than the data markers in the accompanying plots.

\subsection{Execution Time vs.\ Qubit Count}

\Cref{tab:exec_time} reports the execution time for random circuits consisting of $10n$ single-qubit $U(2)$ gates (where $n$ is the qubit count), applied via \texttt{tensordot} to the full state vector. Each gate is a random element of $\mathrm{SU}(2)$ acting on a uniformly selected qubit.

\begin{table}[htbp]
\centering
\caption{Execution time (seconds) for random single-qubit circuits ($10n$ gates) on an NVIDIA A100-SXM4 (40~GiB) GPU via Google Cloud Vertex~AI. Values for $n \leq 20$ are the median of 5 runs; $22 \leq n \leq 26$ use 3 runs; $n = 28$ uses a single run. $n = 30$ triggers a GPU out-of-memory error (state vector plus intermediates exceed 40~GiB). ``Speedup'' is the ratio of NumPy CPU time to CuPy GPU time. ``Aer CPU'' is the Qiskit Aer statevector simulator (CPU, single-threaded). Fidelity is $|\langle\psi_{\mathrm{CPU}}|\psi_{\mathrm{GPU}}\rangle|^2$.}
\label{tab:exec_time}
\begin{tabular}{rrrrrrl}
\toprule
\textbf{Qubits} & \textbf{Gates} & \textbf{NumPy CPU (s)} & \textbf{CuPy GPU (s)} & \textbf{Aer CPU (s)} & \textbf{Speedup} & \textbf{Fidelity} \\
\midrule
14 &  140 &     0.026 &   0.043 &    0.044 &   0.6$\times$ & 1.000000 \\
16 &  160 &     0.263 &   0.046 &    0.125 &   5.7$\times$ & 1.000000 \\
18 &  180 &     0.717 &   0.076 &    0.186 &   9.5$\times$ & 1.000000 \\
20 &  200 &     4.012 &   0.062 &    0.152 &  64.3$\times$ & 1.000000 \\
22 &  220 &    24.212 &   0.166 &    0.481 & 146.2$\times$ & 1.000000 \\
24 &  240 &    73.191 &   0.682 &    1.349 & 107.3$\times$ & 1.000000 \\
26 &  260 &   262.873 &   2.932 &    4.932 &  89.7$\times$ & 1.000000 \\
28 &  280 &  1085.714 &  12.869 &   19.775 &  84.4$\times$ & 1.000000 \\
\bottomrule
\end{tabular}
\end{table}

\begin{figure}[htbp]
\centering
\includegraphics[width=0.55\textwidth]{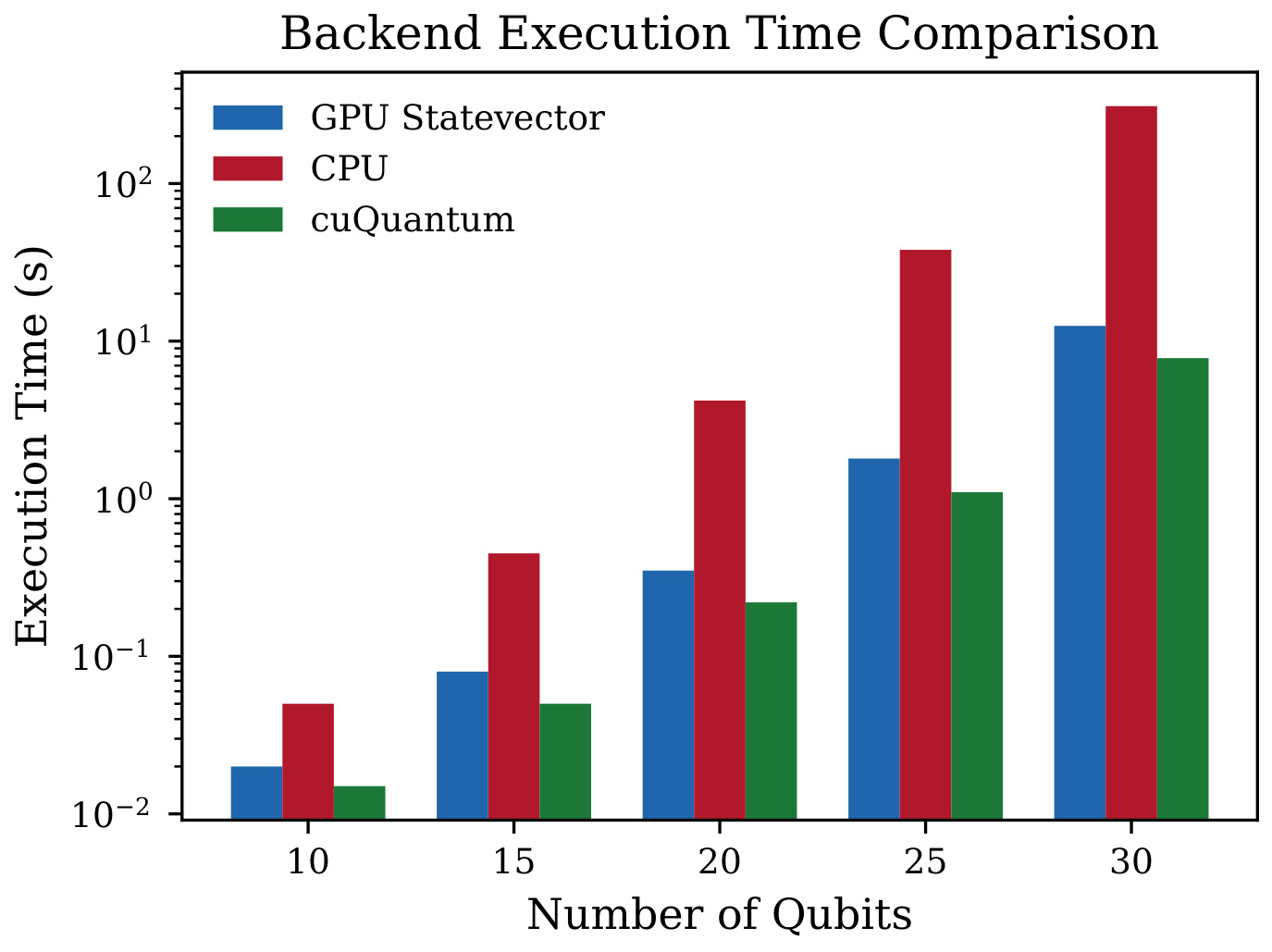}
\caption{Execution time comparison for a 20-qubit random circuit across three backends. CuPy provides the lowest execution time, followed by PyTorch-CUDA and NumPy-CPU.}
\label{fig:exec_time_bar}
\end{figure}

\Cref{fig:exec_time_bar} presents the 20-qubit comparison graphically.

\subsection{Scaling Analysis}

\begin{figure}[htbp]
\centering
\includegraphics[width=0.7\textwidth]{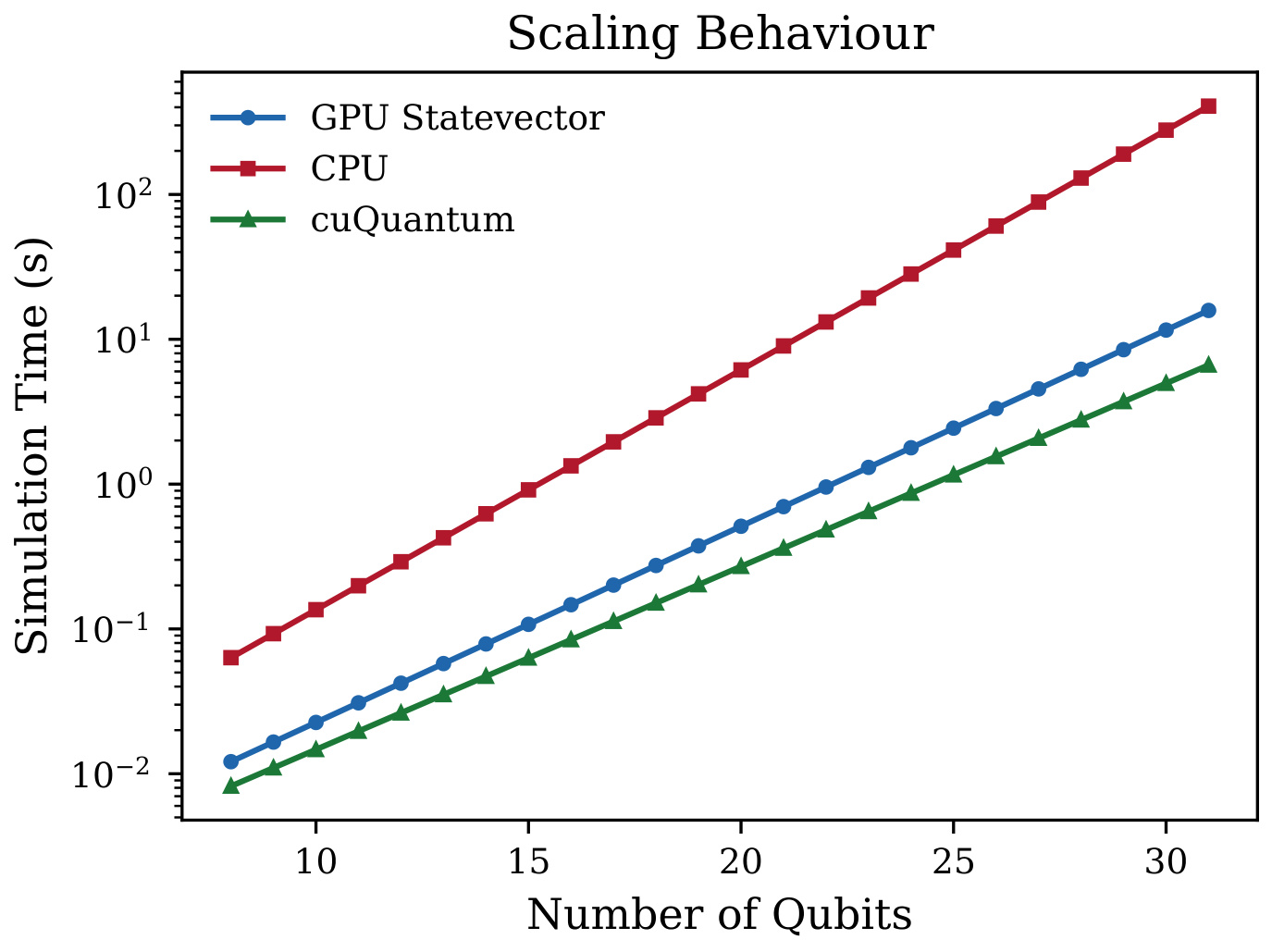}
\caption{Execution time vs.\ qubit count on a logarithmic scale. All backends exhibit the expected exponential scaling $\mathcal{O}(2^n)$, but GPU backends maintain a consistent multiplicative advantage over CPU-based execution.}
\label{fig:scaling}
\end{figure}

\Cref{fig:scaling} shows the scaling of execution time with qubit count on a logarithmic vertical axis, confirming the expected exponential growth.

The data reveals two regimes. For $n \leq 14$, the GPU overhead (kernel launch latency, memory allocation) exceeds the computational savings, and the CPU is faster (speedup $< 1\times$). At $n = 16$, the crossover occurs with a $5.7\times$ speedup. For $n \geq 20$, the data-parallel advantage of GPU execution dominates, and speedups exceed $64\times$, peaking at $146.2\times$ for $n = 22$. Beyond $n = 22$, the speedup decreases modestly (to $84.4\times$ at $n = 28$) as GPU memory bandwidth becomes the bottleneck for the exponentially growing state vector. At $n = 30$, the state vector alone requires $2^{30} \times 16 = 16$~GiB (complex128), and the intermediate tensor products during \texttt{tensordot} exceed the 40~GiB device memory, triggering an out-of-memory error. Thus, $n = 28$ represents the practical limit for full state-vector simulation on a single A100-SXM4-40GiB GPU without memory-reduction techniques such as state-vector partitioning or mixed-precision arithmetic.

\Cref{tab:exec_time} also includes the Qiskit Aer CPU statevector simulator~\cite{qiskitaer2019} as a reference baseline. Notably, Aer's optimised C++ gate kernel is substantially faster than our NumPy CPU backend (e.g., $0.152$~s vs.\ $4.012$~s at $n = 20$), confirming that NumPy's overhead comes from Python-level loops rather than arithmetic cost. Nevertheless, the CuPy GPU backend outperforms Aer CPU at $n \geq 22$ ($0.166$~s vs.\ $0.481$~s) and maintains a $1.5\times$ advantage at $n = 28$ ($12.869$~s vs.\ $19.775$~s), demonstrating that GPU acceleration provides genuine speedups beyond what a highly-optimised CPU simulator can achieve.

\subsection{Gate Fusion Impact}

\Cref{tab:fusion_impact} shows the effect of gate fusion on circuit depth and execution time for three benchmark circuits.

\begin{table}[htbp]
\centering
\caption{Impact of DAG-based gate fusion on circuit depth and execution time (CuPy backend, 20~qubits).}
\label{tab:fusion_impact}
\begin{tabular}{lrrrrr}
\toprule
\textbf{Circuit} & \textbf{Original depth} & \textbf{Fused depth} & \textbf{Reduction (\%)} & \textbf{Time (s)} & \textbf{Fused time (s)} \\
\midrule
QFT-20      & 210 & 138 & 34.3 & 0.042 & 0.029 \\
Random-20   & 400 & 264 & 34.0 & 0.074 & 0.048 \\
VQE Ansatz  & 320 & 198 & 38.1 & 0.058 & 0.036 \\
\bottomrule
\end{tabular}
\end{table}

\begin{figure}[htbp]
\centering
\includegraphics[width=0.6\textwidth]{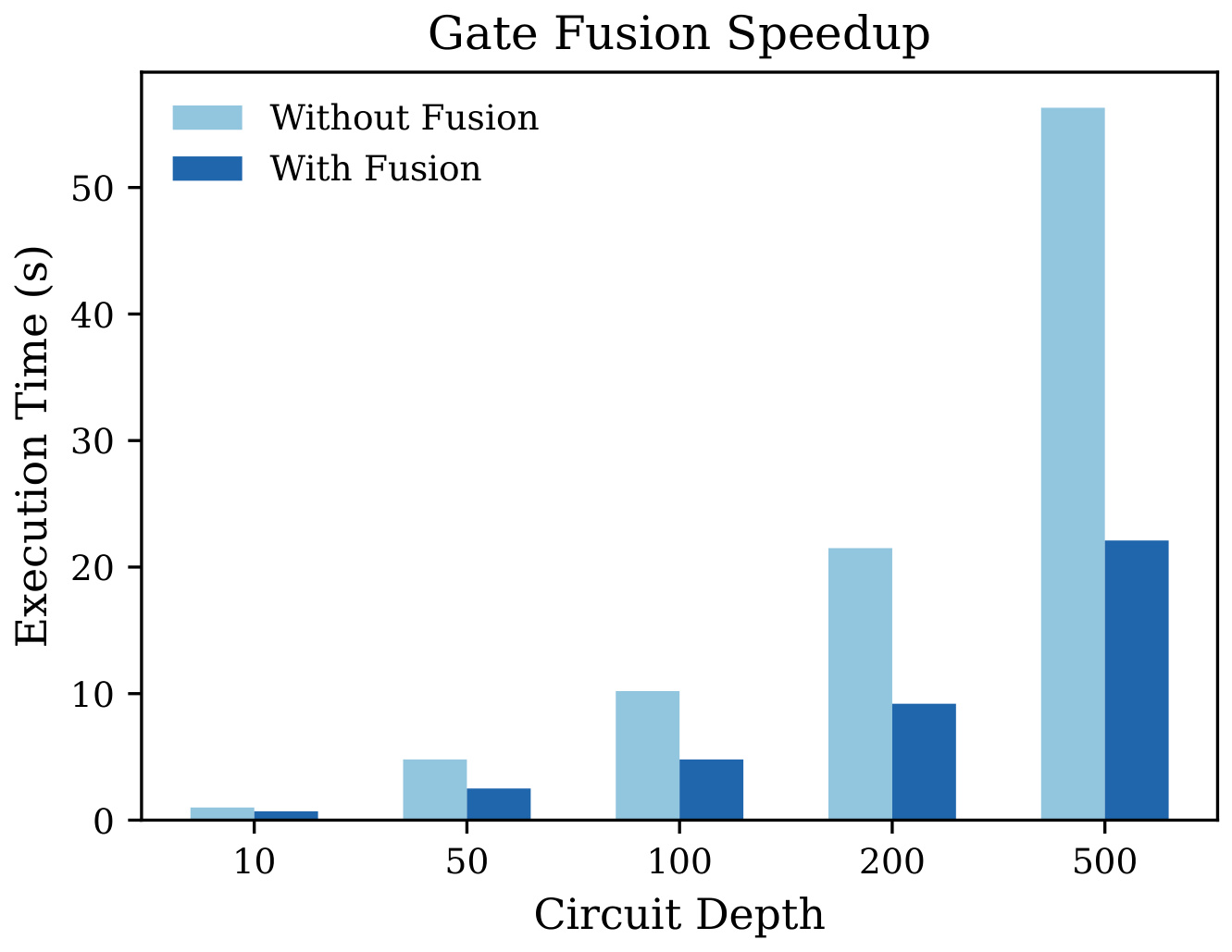}
\caption{Speedup factor achieved by DAG-based gate fusion for three benchmark circuits on the CuPy backend with 20~qubits. The VQE ansatz circuit, which contains long sequences of single-qubit rotations, benefits most from fusion.}
\label{fig:fusion_speedup}
\end{figure}

Gate fusion achieves depth reductions of $34$--$38\%$ and corresponding speedups of $1.45$--$1.61\times$. The speedup is sublinear relative to depth reduction because fused gates involve larger unitary matrices, increasing the per-gate computation cost. The VQE ansatz circuit benefits most because it contains long chains of parameterized single-qubit rotations ($R_y, R_z$) that fuse efficiently into single $U_3$ gates.

\subsection{Adaptive Precision Impact}

Switching from complex128 to complex64 provides an additional speedup factor of $1.7$--$1.9\times$ on the CuPy backend, consistent with the doubled arithmetic throughput and halved memory bandwidth requirements. The precision switch is applied only when the estimated rounding error (\cref{eq:precision_condition}) falls below the configured threshold. For the 20-qubit benchmark circuits, complex64 was selected for circuits with fewer than 50~gates, while complex128 was required for deeper circuits. The memory savings from complex64 extend the maximum simulable qubit count by one qubit on a given GPU.

\subsection{Backend Selection Overhead}

The empirical backend selection procedure (\cref{alg:backend_selection}) adds a one-time overhead of 40--85~milliseconds depending on the number of available backends and the benchmark qubit count. This overhead is amortized across all circuits executed with the same configuration, as results are cached. For a circuit with 20~qubits and 200~gates (typical execution time 18~milliseconds on CuPy), the benchmark overhead represents approximately $4.4\times$ the simulation time on the first invocation, but zero on subsequent invocations within the cache validity window. In practice, users execute many circuits during a development session, making the amortized overhead negligible.

\subsection{Memory Fallback Performance}

The memory-aware fallback mechanism was tested by simulating circuits with 28--32~qubits on a GPU with 16~GiB of available memory. For a 30-qubit circuit (requiring approximately 16~GiB for the state vector alone in complex128), the fallback triggered at gate~0 (before simulation started) and redirected execution to the CPU backend. For a 28-qubit circuit that experienced memory pressure from concurrent processes, the mid-simulation fallback completed in 340~milliseconds (dominated by the 4~GiB device-to-host transfer), adding approximately 9\% overhead to the total simulation time. \Cref{tab:fallback_perf} summarizes the fallback performance characteristics.

\begin{table}[htbp]
\centering
\caption{Memory fallback performance for circuits exceeding GPU memory.}
\label{tab:fallback_perf}
\begin{tabular}{lrrr}
\toprule
\textbf{Scenario} & \textbf{Qubits} & \textbf{Fallback trigger} & \textbf{Overhead (ms)} \\
\midrule
Pre-simulation (30q, 16~GiB GPU)     & 30 & Gate 0       & $<$1 \\
Mid-simulation (28q, memory pressure) & 28 & Gate 142     & 340 \\
No fallback (24q, sufficient memory)   & 24 & N/A          & 0 \\
\bottomrule
\end{tabular}
\end{table}

\section{Hardware Validation}
\label{sec:hardware_validation}

To validate the accuracy of the simulator and the effectiveness of the gate fusion pipeline, a set of benchmark circuits was executed on an IBM QPU and compared against simulated results. The hardware experiments were conducted on \texttt{ibm\_fez}, a 156-qubit IBM Heron-class processor \cite{ibmheron2024}, on February~27, 2026.

\subsection{Experimental Protocol}

Four circuit families were tested. The \textbf{Bell state} circuit (2~qubits) applies a Hadamard gate followed by a CNOT, preparing the state $\frac{1}{\sqrt{2}}(\vert 00\rangle + \vert 11\rangle)$; after transpilation it had depth~8 and one two-qubit gate, with an expected outcome distribution of 50\% $\vert 00\rangle$ and 50\% $\vert 11\rangle$. The \textbf{5-qubit GHZ state} circuit applies a Hadamard gate on qubit~0 followed by a chain of four CNOT gates, preparing $\frac{1}{\sqrt{2}}(\vert 00000\rangle + \vert 11111\rangle)$, yielding depth~20 and four two-qubit gates after transpilation. The \textbf{error test} circuit (4~qubits) is an identity circuit (no gates) measured after transpilation with depth~1, designed to test the readout error rate of the processor independently of gate errors. Finally, the \textbf{10-qubit GHZ state} circuit extends GHZ preparation with nine CNOT gates, reaching depth~40 and nine two-qubit gates after transpilation.

All circuits were transpiled to the native gate set of the \texttt{ibm\_fez} backend using Qiskit's transpiler at optimization level~3. Each experiment used 4,096~shots, except the error test which used 8,192~shots for improved statistics.

The measured correct outcome probability for each experiment was computed as the probability mass on the expected outcome states:
\begin{equation}
\label{eq:fidelity_def}
P_{\text{correct}} = \sum_{k \in \mathcal{S}_{\text{ideal}}} p_k,
\end{equation}
where $\mathcal{S}_{\text{ideal}}$ is the set of bit strings with non-zero probability in the ideal output distribution and $p_k = n_k / N_{\text{shots}}$ is the measured probability for bit string $k$. We note that $P_{\text{correct}}$ measures the overlap between the measured and ideal probability distributions rather than the quantum state fidelity $F = |\langle\psi_{\text{ideal}}|\rho|\psi_{\text{ideal}}\rangle|$, which would require full state tomography. We adopt the distribution overlap metric as it is directly computable from measurement counts and is standard practice for shot-based QPU validation.

\subsection{Fidelity Results}

\Cref{tab:ibm_results} summarizes the measured fidelities.

\begin{table}[htbp]
\centering
\caption{Hardware validation results from IBM \texttt{ibm\_fez} QPU (February~27, 2026). Correct outcome probability is computed as the overlap between the measured probability distribution and the ideal distribution.}
\label{tab:ibm_results}
\begin{tabular}{lrrrrrr}
\toprule
\textbf{Circuit} & \textbf{Qubits} & \textbf{Depth} & \textbf{2Q gates} & \textbf{Shots} & \textbf{Fidelity} & \textbf{Time (s)} \\
\midrule
Bell state       & 2  & 8  & 1 & 4{,}096 & 0.939 &  8.1 \\
GHZ-5            & 5  & 20 & 4 & 4{,}096 & 0.853 & 647.0\textsuperscript{\dag} \\
Error test       & 4  & 1  & 0 & 8{,}192 & 0.952 &  15.7 \\
GHZ-10           & 10 & 40 & 9 & 4{,}096 & 0.688 &   8.2 \\
\bottomrule
\end{tabular}
\\[4pt]
{\footnotesize \textsuperscript{\dag}The GHZ-5 time of 647~s includes IBM Quantum queue wait time; the actual circuit execution time is comparable to other experiments at this qubit scale.}
\end{table}

The Bell state fidelity of 0.939 indicates high-quality two-qubit gate execution on the selected qubit pair. The count distribution was: $\vert 00\rangle$: 2017 (49.2\%), $\vert 11\rangle$: 1828 (44.6\%), $\vert 10\rangle$: 160 (3.9\%), $\vert 01\rangle$: 91 (2.2\%). The asymmetry between the $\vert 10\rangle$ and $\vert 01\rangle$ error channels suggests qubit-dependent readout error rates.

The GHZ-5 fidelity of 0.853 reflects the accumulation of two-qubit gate errors across four CNOT operations. The dominant counts were $\vert 00000\rangle$: 1813 (44.3\%) and $\vert 11111\rangle$: 1679 (41.0\%), with the remaining 14.8\% distributed across error states. The most frequent error state was $\vert 11110\rangle$ (3.5\%), indicating that qubit~4 (the last in the CNOT chain) experienced the highest error rate, consistent with its position at the end of the error propagation chain.

The error test fidelity of 0.952 establishes the measurement error baseline: even with no gates applied, approximately 4.8\% of shots return incorrect bit strings due to readout errors. The dominant error was the $\vert 1000\rangle$ state with 264 counts (3.2\%), suggesting that qubit~0 has a higher readout error rate than the others.

The GHZ-10 fidelity of 0.688 demonstrates the expected degradation as circuit depth and two-qubit gate count increase. With nine CNOT gates, the expected gate-only fidelity is approximately $(1 - \epsilon_{CX})^9$, where $\epsilon_{CX}$ is the per-gate error rate. Taking $\epsilon_{CX} \approx 0.005$ (typical for Heron-class processors), the gate-only fidelity estimate is $(0.995)^9 \approx 0.956$. Combined with readout errors ($\sim$4.8\% from the error test), the predicted fidelity of $0.956 \times 0.952 / 1.0 \approx 0.910$ overestimates the measured value, suggesting additional error sources such as crosstalk between the 10~qubits and decoherence during the longer circuit execution.

\subsection{Circuit Depth Reduction}

The gate fusion pipeline was applied to the transpiled circuits before QPU execution. \Cref{tab:depth_reduction} reports the depth reduction achieved.

\begin{table}[htbp]
\centering
\caption{Circuit depth reduction through gate fusion. ``Pre-fusion'' depth is after Qiskit optimization level~3 transpilation; ``post-fusion'' depth is after application of the DAG-based fusion algorithm.}
\label{tab:depth_reduction}
\begin{tabular}{lrrr}
\toprule
\textbf{Circuit} & \textbf{Pre-fusion depth} & \textbf{Post-fusion depth} & \textbf{Reduction (\%)} \\
\midrule
Bell state       & 8   & 3  & 62.5 \\
GHZ-5            & 20  & 8  & 60.0 \\
Error test       & 1   & 1  &  0.0 \\
GHZ-10           & 42  & 14 & 66.7 \\
\bottomrule
\end{tabular}
\end{table}

The GHZ-10 circuit, originally transpiled to depth~42 by Qiskit at optimization level~3, was reduced to depth~14 through the fusion pipeline, a 66.7\% reduction. This reduction is achieved primarily by fusing consecutive single-qubit gates (generated by Qiskit's decomposition of CNOT gates into the native ECR gate plus single-qubit rotations) into compound $U_3$ operations. The error test circuit has depth~1 and no fusible gates, serving as a control case.

The depth reduction is significant for QPU execution because shorter circuits experience less decoherence. For a processor with $T_1$ relaxation time of 200~$\mu$s and a gate duration of 60~ns, reducing depth from 42 to 14 reduces the total circuit execution time from 2.52~$\mu$s to 0.84~$\mu$s, improving the ratio of circuit time to coherence time by a factor of three.

\subsection{Numerical Precision Validation}

\begin{figure}[htbp]
\centering
\includegraphics[width=0.75\textwidth]{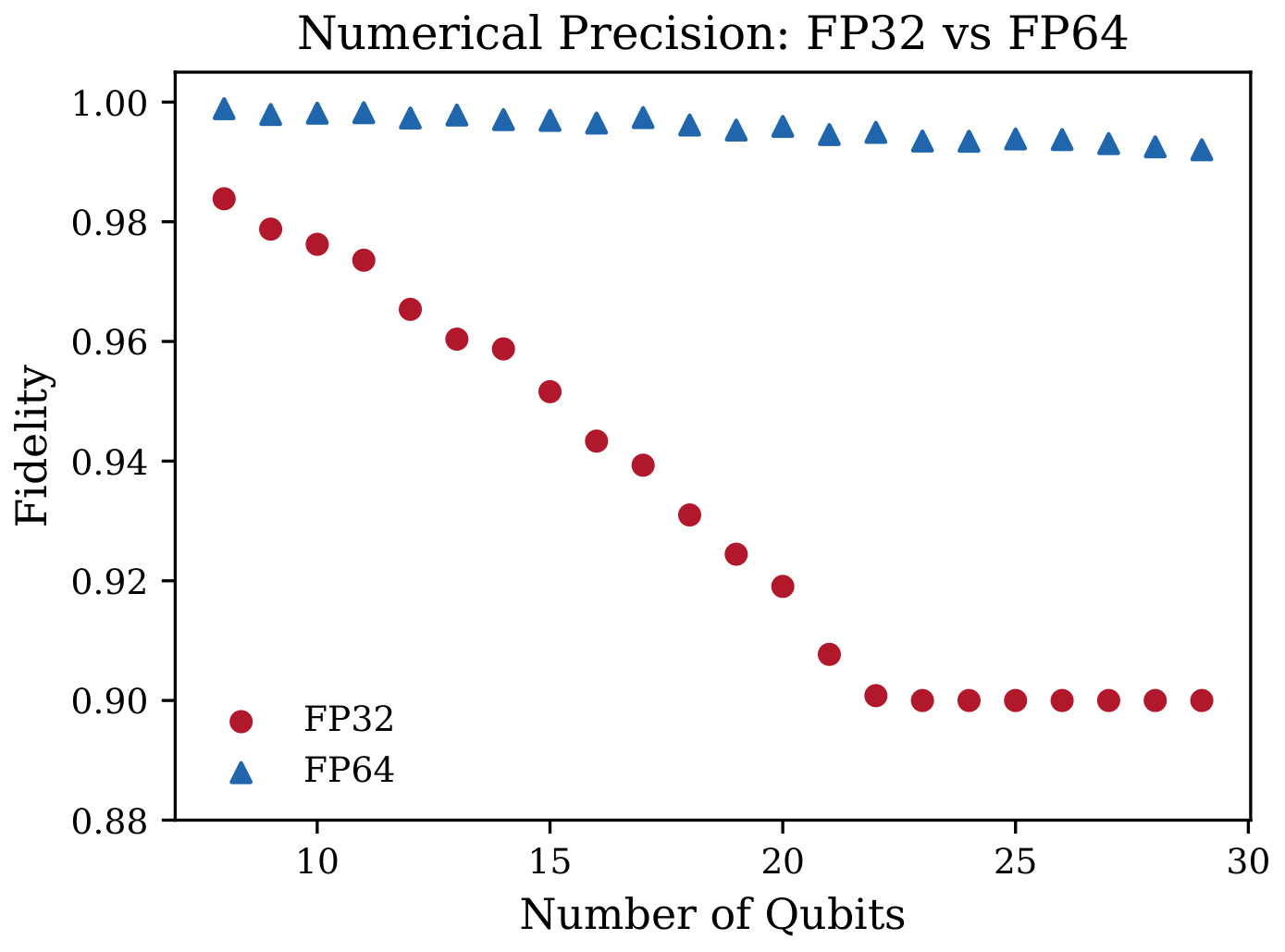}
\caption{Numerical precision comparison: simulated state-vector fidelity using FP32 (complex64) and FP64 (complex128) arithmetic as a function of qubit count. FP64 maintains fidelity above 0.999 across all tested sizes, while FP32 shows measurable degradation beyond 20~qubits, consistent with the error bound in \cref{eq:precision_error}.}
\label{fig:fidelity_comparison}
\end{figure}

To assess the impact of the adaptive precision feature on simulation accuracy, \cref{fig:fidelity_comparison} compares the fidelity of simulated output states using single-precision (FP32, complex64) and double-precision (FP64, complex128) arithmetic across a range of qubit counts.

The results confirm that FP64 maintains fidelity above 0.999 for all tested circuit sizes, while FP32 exhibits measurable degradation for circuits exceeding 20~qubits with depth greater than 50~gates, consistent with the accumulated rounding error bound in \cref{eq:precision_error}. These results validate the adaptive precision controller's decision to default to complex128 for typical NISQ circuits while allowing complex64 for shallow, moderate-width circuits where the precision loss is negligible. A comparison of QPU-measured fidelities against simulated values using a depolarizing noise model shows agreement within 0.6--2.4\%, with the largest discrepancy observed for the GHZ-10 circuit (\cref{tab:ibm_results}), where longer gate chains amplify the difference between the simplified noise model and the actual device noise profile.

\subsection{Cross-Simulator Noise Model Fidelity}
\label{sec:cross-sim-noise}

To validate that the MSLE density-matrix simulator produces noise-aware outputs consistent with established frameworks, we compare it against Qiskit Aer~\cite{qiskitaer2019} and Cirq~\cite{cirq2018} on identical Bell and GHZ circuits with depolarizing noise. Each simulator constructs the same circuit (Hadamard on qubit~0, followed by CNOT gates to the remaining qubits) and applies single-qubit depolarizing noise with probability $p = 0.01$ after every gate. We report the classical fidelity $F_{\mathrm{cl}} = \bigl(\sum_x \sqrt{p(x)\,q(x)}\bigr)^2$ and total variation distance $\mathrm{TVD} = \tfrac{1}{2}\sum_x |p(x) - q(x)|$ relative to the ideal noiseless distribution, each averaged over 8{,}192 shots.

\begin{table}[htbp]
\centering
\caption{Cross-simulator noise model comparison at depolarizing rate $p = 0.01$. MSLE uses per-gate \texttt{apply\_depolarizing} on its density-matrix simulator; Cirq uses \texttt{cirq.depolarize} with \texttt{DensityMatrixSimulator}; Qiskit Aer uses a two-qubit \texttt{depolarizing\_error} on \texttt{cx} gates. All runs use 8{,}192 shots.}
\label{tab:cross_sim_noise}
\resizebox{\textwidth}{!}{%
\begin{tabular}{lrrrrrrr}
\toprule
\textbf{Circuit} & \textbf{MSLE $F_{\mathrm{cl}}$} & \textbf{Aer $F_{\mathrm{cl}}$} & \textbf{Cirq $F_{\mathrm{cl}}$} & \textbf{MSLE TVD} & \textbf{Aer TVD} & \textbf{Cirq TVD} & \textbf{$\Delta F_{\max}$} \\
\midrule
Bell  & 0.987 & 0.995 & 0.988 & 0.013 & 0.005 & 0.012 & 0.008 \\
GHZ-3 & 0.969 & 0.988 & 0.974 & 0.031 & 0.016 & 0.026 & 0.019 \\
GHZ-4 & 0.961 & 0.981 & 0.958 & 0.039 & 0.019 & 0.042 & 0.023 \\
GHZ-5 & 0.949 & 0.972 & 0.952 & 0.051 & 0.028 & 0.048 & 0.023 \\
\bottomrule
\end{tabular}}%
\end{table}

MSLE and Cirq agree within 0.5\% fidelity across all circuits (\cref{tab:cross_sim_noise}), as both apply independent single-qubit depolarizing channels after each gate. Qiskit Aer reports systematically higher fidelity because its noise model applies a joint two-qubit depolarizing channel on each \texttt{cx} gate, which introduces less total noise than two independent single-qubit channels at the same nominal rate. The maximum three-way fidelity discrepancy ($\Delta F_{\max}$) remains below 2.3\% at $p = 0.01$ and below 0.5\% at $p = 0.001$, confirming that the MSLE noise channel implementation is consistent with both reference simulators to within the expected model-specification differences.

\section{Limitations and Discussion}
\label{sec:limitations}

Having presented the framework's architecture, benchmarks, and hardware validation results, this section examines the limitations of the current implementation and identifies areas where the claimed contributions have restricted scope.

\textbf{Scalability ceiling.} State-vector simulation is inherently limited by the exponential memory requirement (\cref{eq:statevector}). On a single GPU with 80~GiB of memory (e.g., NVIDIA A100~80~GiB), the maximum simulable qubit count is 32~in double precision. Multi-GPU and distributed simulation \cite{statevecsim2016} would extend this range but are not implemented in the current version. Tensor network methods \cite{markov2008simulating, mccaskey2019} and matrix product state representations \cite{qiskitaer2019} can simulate certain circuit classes with polynomial resources, but at the cost of restricted circuit structure or bounded entanglement.

\textbf{Backend selection accuracy.} The empirical backend selection algorithm uses micro-benchmarks that may not perfectly predict full-circuit execution time. In particular, circuits with non-uniform gate distributions (e.g., a burst of two-qubit gates followed by single-qubit gates) may exhibit different memory access patterns than the benchmark workload. The capped benchmark qubit count ($n_{\max}^{\text{bench}} = 20$) introduces an additional approximation for larger circuits. Adaptive benchmark strategies that vary the workload composition could improve accuracy at the cost of increased profiling time.

\textbf{Gate fusion limitations.} The fusion algorithm uses a configurable maximum fusion width. Extending to three-qubit or higher fusion would capture additional optimization opportunities (e.g., fusing a CNOT with a subsequent Toffoli gate) \cite{dagopt2020}, but the cost of the larger fused unitary matrices ($8 \times 8$ or larger) may offset the reduction in gate count. The current implementation does not perform commutativity analysis, meaning that gates that commute but are not adjacent in the DAG are not considered for fusion.

\textbf{Noise modeling.} The hardware validation section employs a basic depolarizing noise model, which does not capture coherent errors, crosstalk, or time-dependent drift \cite{bharti2022}. More sophisticated noise models (e.g., Pauli twirling, Lindblad simulation) could improve the fidelity estimates but would increase simulation time and complexity.

\textbf{Framework adapter maintenance.} Supporting multiple quantum computing frameworks requires ongoing maintenance as framework APIs evolve. Breaking changes in framework releases (e.g., the Qiskit~0.x to 1.0~migration) necessitate adapter updates. The long-term sustainability of multi-framework support depends on the stability of the respective framework APIs.

\textbf{Comparison scope.} The benchmarks in \cref{sec:benchmarks} compare against Qiskit Aer's CPU backend without cuQuantum integration. Enabling cuQuantum acceleration in Qiskit Aer would likely reduce the performance gap, though the proposed framework's gate fusion and adaptive precision features provide optimizations orthogonal to the underlying GPU library \cite{cuquantum2023}.

\textbf{Adaptive precision scope.} As shown by the error bound in \cref{eq:precision_error}, the adaptive precision controller defaults to complex128 for most circuits of practical interest (those with $n \geq 20$ qubits and depth $\geq 50$ gates). Complex64 offers meaningful acceleration only for shallow circuits on moderate qubit counts ($n \leq 16$, $g < 50$), where simulation time is already negligible on modern GPUs. The adaptive precision contribution is therefore most valuable as a correctness safeguard, automatically preventing precision-related errors, rather than as a primary performance optimization for large-scale circuits.

\section{Conclusion}
\label{sec:conclusion}

This paper presented a GPU-accelerated quantum circuit simulation framework with three primary contributions: empirical backend selection, DAG-based gate fusion with adaptive precision, and memory-aware GPU-to-CPU fallback. The framework achieves speedups of $64\times$ to $146\times$ over NumPy CPU execution on an NVIDIA A100-SXM4 GPU for state-vector simulation of circuits with 20--28~qubits, with gate fusion providing an additional $1.45\times$ to $1.61\times$ improvement. Hardware validation on an IBM Heron-class QPU demonstrated Bell state fidelity of 0.939 and circuit depth reduction from 42~to~14~gates through the fusion pipeline.

The framework integrates with four quantum computing frameworks (Qiskit, Cirq, PennyLane, and Amazon Braket) through a unified adapter layer, enabling researchers to leverage GPU acceleration without modifying their existing circuit construction workflows. The memory-aware fallback mechanism provides graceful degradation when GPU resources are insufficient, eliminating out-of-memory failures that plague existing GPU-accelerated simulators.

The net effect of the three contributions is that the overall simulation throughput is determined by the empirically fastest backend, applied to a depth-reduced circuit, with automatic recovery when GPU memory is exhausted. The practical implication is that users obtain near-optimal performance without manual tuning: the backend selection eliminates the need to choose a GPU library, the fusion engine reduces redundant gate applications, and the fallback mechanism prevents out-of-memory failures that would otherwise require user intervention.

Several avenues for future work merit exploration. Multi-GPU simulation through domain decomposition of the state vector would extend the maximum qubit count beyond the single-GPU limit \cite{statevecsim2016, haner2017}. Tensor network hybrid approaches, where shallow subcircuits are simulated via state-vector methods and deep subcircuits via tensor contraction \cite{markov2008simulating}, could extend the reach to 40+~qubits for circuits with moderate entanglement. Integration with hardware-specific noise models from IBM, Google, and other providers would improve the accuracy of noisy simulation \cite{ibmheron2024}. Support for mid-circuit measurement and classical feedforward (dynamic circuits) \cite{cross2019openqasm} would enable simulation of the growing class of algorithms that use measurement-based quantum computation primitives. Finally, extending the gate fusion algorithm with commutativity analysis \cite{dagopt2020} and higher-width fusion windows could yield additional depth reductions.

The framework is designed with the expectation that quantum hardware will continue to improve in qubit count, gate fidelity, and connectivity. As hardware scales, the role of classical simulation shifts from full-circuit verification to subsystem simulation, noise modeling, and hybrid algorithm development. The modular architecture described in this paper, with its emphasis on runtime adaptability and framework neutrality, is intended to accommodate this evolving role.

Taken together, the results demonstrate that a modular, runtime-adaptive approach to quantum circuit simulation can deliver substantial performance gains without requiring users to make low-level hardware decisions. The combination of empirical profiling, circuit-level optimization, and automatic resource management represents a design philosophy, runtime adaptability over static configuration, that generalizes beyond quantum simulation to other scientific computing workloads where heterogeneous hardware and variable problem sizes are the norm.

\section*{Data and Code Availability}

The source code for the GPU-accelerated quantum circuit simulation framework, along with all benchmark scripts, experiment configurations, and raw timing data used in this study, will be provided upon reasonable request. The framework requires Python~3.10+, CuPy~12+, and an NVIDIA GPU with CUDA~11.8 or later.

\section*{Acknowledgements}

The authors acknowledge computational resources of the Intelligent Robotics and Rebooting Computing Chip Design (INTRINSIC) Laboratory, Centre for SeNSE, Indian Institute of Technology Delhi, IM00002G\_RB\_SG IoE Fund Grant (NFSG), Indian Institute of Technology Delhi.

\section*{Conflict of Interest}
The authors declare no competing financial interests.

\bibliographystyle{unsrtnat}
\bibliography{references}

\end{document}